\documentclass[pra,onecolumn,preprint,showpacs,groupedaddress,superscriptaddress,nofootinbib,floatfix,preprintnumbers,longbibliography,aps]{revtex4-2}

\usepackage{graphicx}% Include figure files
\usepackage{dcolumn}% Align table columns on decimal point
\usepackage{bm}% bold math
\usepackage{amsmath}
\usepackage{amssymb}
\usepackage{gensymb}
\usepackage{braket}
\usepackage{xcolor}
\newcommand{\vbA}{\bm{A}}
\newcommand{\vbr}{\bm{r}}

\newcommand{\vbK}{\bm{K}}
\newcommand{\vbk}{\bm{k}}
\newcommand{\vbB}{\bm{B}}
\newcommand{\vbE}{\bm{E}}
\newcommand{\infi}{\ensuremath{\mathrm{if}}}
\newcommand{\vbd}{\bm{d}}
\newcommand{\vbpi}{\bm{\Pi}}
\newcommand{\vbv}{\bm{v}}
\newcommand{\eref}[1]{(\ref{#1})}

\newcommand{\Eref}[1]{Equation (\ref{#1})}

\def\cases#1{%
     \left\{\,\vcenter{\def\\{\cr}\normalbaselines\openup1\jot\m@th%
     \ialign{\strut$\displaystyle{##}\hfil$&\tqs
     \rm##\hfil\crcr#1\crcr}}\right.}%
\newcommand{\tqs}{\hspace*{25pt}}

\newcommand{\rme}{\mathrm{e}}

\newcommand{\rmd}{\mathrm{d}}
\newcommand{\br}{\hline}
\newcommand{\mr}{\hline}
\begin{document}
%\preprint{APS/123-QED}

\title{Nondipole circularly polarized laser-assisted photoelectron emission}% Force line breaks with \\

\author{R. Della Picca}
\email{renata@cab.cnea.gov.ar}
\affiliation{Centro At\'omico Bariloche (CNEA),  CONICET and Instituto Balseiro (UNCuyo), 8400 Bariloche, Argentina}

\author{J. M. Randazzo}
\affiliation{Centro At\'omico Bariloche (CNEA),  CONICET and Instituto Balseiro (UNCuyo), 8400 Bariloche, Argentina}

\author{S. D. L\'opez}
\affiliation{Instituto de Investigaciones en Energía No Convencional - INENCO (UNSa - CONICET), Av. Bolivia 5150, 4400, Salta Capital, Argentina}

\author{M. F. Ciappina}
\email{marcelo.ciappina@gtiit.edu.cn}
\affiliation{Department of Physics, Guangdong Technion - Israel Institute of Technology, 241 Daxue Road, Shantou, Guangdong, China, 515063}
\affiliation{Technion - Israel Institute of Technology, Haifa, 32000, Israel}
\affiliation{Guangdong Provincial Key Laboratory of Materials and Technologies for Energy Conversion, Guangdong Technion - Israel Institute of Technology, 241 Daxue Road, Shantou, Guangdong, China, 515063}

\author{D. G. Arb\'o }
\affiliation{Institute for Astronomy and Space Physics - IAFE (UBA-Conicet), C1428GA Buenos Aires, Argentina}
\affiliation{Universidad de Buenos Aires - Facultad de Ciencias Exactas y Naturales and Ciclo B\'asico Com\'un,  C1428EGA Buenos Aires, Argentina}

\date{\today}% It is always \today, today,

\begin{abstract}
We theoretically study atomic laser-assisted photoelectric emission (LAPE) beyond the electric dipole approximation. 
We present a theoretical description for first-order nondipole corrections ($O(c^{-1})$ where $c$ is the speed of light) to the nonrelativistic description of the laser-atom interaction for a strong circularly polarized infrared (IR) laser field combined with a train of extreme-ultraviolet (XUV) laser pulses. 
We investigate the photoelectron momentum distribution (PMD) as the product of two main contributions: the intra- and interpulse factors. 
Whereas the interpulse factor gives rise to a sideband pattern with a shift opposite to the IR beam propagation direction, the intrapulse factor forms an angular streaking pattern following the IR time-dependent polarization direction. 
We explore the transition of the PMD from the dipole to the nondipole framework, showing the gradual break of the forward-backward symmetry as the laser parameters are varied. 
Furthermore, we find non-zero contributions in dipole forbidden directions independent of the IR polarization state, wherein Cooper-\textit{like} minima are observed. 
Our work lays a theoretical foundation for understanding time-resolved nondipole LAPE in cutting-edge ultrafast experiments.
\end{abstract}

%\keywords{Laser-assisted, photoionization, strong field, nondipole, angular streaking,intercycle, intracycle }%Use showkeys class option if keyword
                    
\maketitle
\newpage
%\tableofcontents

%-------------------------------------------------------
\section{Introduction}
%-------------------------------------------------------

In recent years, there has been increasing interest in exploring strong laser-matter interactions beyond the widely used electric dipole approximation. 
In this context, the assumption of a spatially homogeneous electric field becomes invalid when the laser wavelength is shorter than the size of the atomic system, defining a high-frequency limit. 
However, the opposite limit --very long wavelengths or low frequencies-- is rarely discussed.
In 2014, Reiss criticized the tunneling model at low frequencies \cite{Reiss14}. According to Reiss, the atomic, molecular, and optical community became accustomed to approximate lasers as quasi-static electric (QSE) fields, which can oscillate in time but it does not propagate. 
Instead, he pointed out that true traveling plane waves are \textit{transverse fields} and compared differences between QSE and laser fields. 
Considering the electron-laser interaction in the context of transverse fields, he argues that relativistic, radiation pressure, and magnetic displacement effects delimit the dipole approximation, establishing an enlightening \textit{ ``dipole oasis"} in a frequency-intensity map (see Fig. \ref{fig1:tipoReiss}). 
Outside this region, a nondipole approach needs to be considered.

More recently, Maurer and Keller \cite{Maurer21} expanded on these ideas by providing a comprehensive review of ionization processes in the nondipole regime where a fully relativistic approach is not required. 
They described that during ionization and specifically at longer wavelengths (for example, into the mid-infrared spectral region) or higher intensities, the photoelectron acquires sufficient momentum such that the magnetic component of the laser field (considered negligible within the dipole approximation) plays a significant role through the full Lorentz force. 
Not only is the photoelectron driven by the oscillating electric field but also it experiences a drift in the laser propagation direction due to the influence of the magnetic field. 
This displacement can be estimated by the amplitude of a ``figure-8" motion in the framewrok of the accelerated electron in the direction of laser propagation \cite{Reiss14, Maurer21}:
\begin{equation}
\beta_0 = \frac{U_p}{2 m\omega c} ,
\end{equation}
where $c$ is the speed of light and the ponderomotive potential $U_p=(F_{L0}/2\omega)^2$ depends on the square of the ratio between the laser electric field amplitude and frequency. 
The equation $\beta_0$ equal to the Bohr radius delimits the validity of the dipole approximation for very long wavelengths or small frequencies (Fig.~\ref{fig1:tipoReiss}).
An immediate consequence of this electron trajectory drift is its impact on rescattering processes, i.e., high-above-threshold ionization (HATI) and high-order harmonic generation (HHG). 
Both processes, based on the photoelectron recollision with the parent ion, show a decrease in their probability to occur within the nondipole regime \cite{Maurer21,Lin22, Kahvedzic22,Hartung19,Minneker2022}.
 
Another notable nondipole issue is the linear photon momentum transfer \cite{Maurer21}. 
Since the laser photon carries energy and momentum, conservation rules require a transfer of both quantities to the atomic system.
Meanwhile, energy transfer is well understood within the dipole approximation, the photon momentum transfer remains not fully scrutinized yet. 
This fact has motivated a large production of theoretical and experimental research works \cite{Ilchen2018,Willenberg19,Hartung19,Jensen20,Ni2020,Lin22,Forre2022}, just to cite a few.
An observed relevant feature, precisely due to the photon momentum transfer, is the existence of a forward-backward photoemission asymmetry in the propagation direction.
It has also been shown that the acquired momentum is shared between the electron and the ion in a complex way \cite{Maurer21}. From an experimental perspective, investigating this phenomenon requires measuring the electron and ion momenta in coincidence. From a theoretical standpoint, it demands a highly realistic description of the electron-ion interaction~\cite{Willenberg19,Lin22}. 
These difficulties keep the topic open and under current discussion. 
In fact, Maurer and Keller \cite{Maurer21} have suggested that more studies on photon momentum transfer would be desirable, especially those with attosecond temporal resolution, so that they could provide deeper insight into the underlying mechanism of the electron-ion momentum sharing.

We can mention two typical schemes for attosecond time-resolved measurements: The attoclock and the streak-camera methods (see \cite{Eckle2008,Pfeiffer2012,Goulielmakis2004,Goulielmakis2008,Drescher2005} and references therein). 
The latter utilizes the ionization with synchronized low and high frequency pulses: A typically extreme ultraviolet (XUV) attosecond pulse ionizes the target generating an electron wavepacket in the continuum in the presence of a typically infrared or near-infrared (IR/NIR) intense laser field.
This is a type of laser-assisted photoionization (LAPE) process.
LAPE processes have a long record of studies within the dipole approximation: From the first identification of the sideband peaks \cite{Veniard95} to the present (see, for example, \cite{Romanov2024,DellaPicca2020} and references therein).
Nevertheless, LAPE \textit{beyond} the dipole approximation is a rather unexplored topic. 
To the best of our knowledge, there are only a few theoretical studies on this topic: Jensen \textit{et al.} \cite{Jensen20} have presented a nondipole Hamiltonian from which a first theoretical framework for streaking could be obtained. 
We have also recently focused on nondipole LAPE in the sideband regime for linearly polarized light \cite{DellaPicca2023}. 
Liao \textit{et al.} also investigated nondipole effects in the RABBIT protocol \cite{Liao2024}. 
In addition, more recently, Liang \textit{et al.} \cite{Liang2024}
have measured streaking patterns of helium atoms due to a XUV train pulse phase-locked with an intense IR laser field, resolving the forward-backward asymmetry of the photoelectron yields with attosecond resolution.
%----
% https://arxiv.org/pdf/2210.15856   <--- y este? qué onda?
%----

It is worth highlighting that nondipole LAPE in the streaking regime enables time-resolved measurement schemes, extending attosecond spectroscopy into the nondipole domain (as a recent experiment see e.g.~\cite{Liang2024}).
The aim of this work is to develop a comprehensive theoretical model of LAPE processes to enhance the understanding of time-resolved photoelectron patterns during the transition from dipole to nondipole behaviors. 
We investigate on the case of an attosecond pulse train assisted by a circularly polarized infrared laser, focusing on the streaking spectra as a function of the ionization time.

The paper is organized as follows. 
 In Sec. \ref{theory}, we briefly resume the leading-order nondipole SFA theory of LAPE processes and analyze the properties of the photoelectron momentum distribution (PMD) for a circularly polarized probe IR field and a pump XUV pulse train.
In Sec. \ref{Results}, we analyze the main features and behavior of the PMD as the laser parameters are varied, transitioning from the dipole regime to the nondipole regime.
Concluding remarks are presented in Sec.~\ref{conclusions}.
Atomic units are used throughout the paper, except where otherwise stated.

%------------------------------------------------
\section{Theory} \label{theory}
%------------------------------------------------
\subsection{First order nondipole electromagnetic fields}
%------------------------------------------------

Let us model the laser field, typically with a wavelength in the infrared or near-infrared range, as a traveling electromagnetic (EM) wave described by a monochromatic plane wave with wave vector $\vbK_L  = K_L \hat{z}$ (where $\hat{z}$ is chosen as the propagation direction). 
Starting from the vector potential  $\vbA$ and assuming a null scalar potential, the electric and magnetic fields can be obtained as
 \begin{equation}
     \vbE(\vbr,t) = -\frac{\partial}{\partial t} \vbA(\eta) \qquad \textrm{ and }  \qquad 
     \vbB(\vbr,t) = \nabla \times \vbA(\eta) ,
 \end{equation}
respectively, where the vector potential depends on time and position explicitly through the dimensionless parameter $\eta= \omega t -\vbK_L\cdot\vbr$, where $\omega = c K_L $ is the frequency of the field. 

We are interested in the nondipole and not fully relativistic effects on the laser-atom interaction. 
To explore this, we consider an inhomogeneous vector potential, expressed to the lowest order in $1/c$. 
Specifically, a first order (in $1/c$) Taylor series expansion of the vector potential around $\vbr=0$ yields
\begin{eqnarray}
\vbA(\eta) 
&\simeq& \vbA(\eta)\big|_{\vbr=0} + \sum_{i=1}^3 \frac{\partial}{\partial x_i} \vbA(\eta) \big|_{\vbr=0}\, x_i  \nonumber \\
&\simeq & \vbA_0(t) + \frac{\hat{z}  \cdot \vbr }{c} \vbE_0(t)  , \label{Anodip}
\end{eqnarray}
where the subscript $0$ indicates evaluation of the quantities at the origin: $\vbA_0(t)=\vbA(\eta\big|_{\vbr=0})$ and  $\vbE_0(t)=\vbE(\vbr=0,t) = - \frac{\partial}{\partial t} \vbA_0(t)  $.
Considering the vector potential from Eq.~(\ref{Anodip}), the electric and magnetic fields become
\begin{subequations}\label{EM2}
\begin{eqnarray}
    \vbE(\vbr,t) &\simeq& \vbE_0(t)  \\
    \vbB(\vbr,t) &\simeq& \frac{1}{c} \hat{z}\times \vbE_0(t) . 
\end{eqnarray}
\end{subequations}

Since the interaction Hamiltonian between a free electron ($e$) and the EM field is $ - i \vbA\cdot\nabla + \frac{1}{2} \vbA^2 $, the gauge transformation $\Psi^L = \exp(-i \vbA_0\cdot \vbr) \Psi $ allows us to identify an alternative interaction Hamiltonian of the form
\begin{equation}
   H_{\mathrm{eEM}}^L = \vbE_0(t) \cdot \left[ \vbr -i \frac{(\hat{z} \cdot \vbr)}{c}  \nabla   \right].
\end{equation}
It can be shown that the so-called nondipole Gordon-Volkov wavefunction in the length gauge (see  \cite{Kylstra01} and Eq.~(2.199) of \cite{Joachain12})
\begin{eqnarray}
 \chi^{V_{\mathrm{ND}}}(\vbr,t) &=& 
 (2\pi)^{-3/2} \exp \left[ i\, \vbpi(\vbk,t) \cdot \vbr \right]\;\exp{ \frac{i}{2} \int_{t}^{\infty} \vbpi^2(\vbk,t^\prime) \rmd t^\prime }, \label{volkovNONdipo}
\end{eqnarray}
verifies the following Schr\"odinger equation
\begin{equation}
    i ~ \frac{\partial}{\partial t}  \chi^{V_{\mathrm{ND}}}   = \left[-\frac{\nabla^2}{2} +   H_\mathrm{eEM}^L \right]  \chi^{V_{\mathrm{ND}}} + O(1/c^2).
\end{equation}
describing an electron with momentum $\vbk$ moving in an intense laser field, up to order $c^{-1}$. 
In Eq.~(\ref{volkovNONdipo}), we have introduced the ``nondipole effective momentum" $\vbpi$ as
\begin{eqnarray}
\vbpi(\vbk,t) &=& \vbk + \vbA_0(t) + \big[ \vbk \cdot \vbA_0(t) + \frac{1}{2} \vbA_0^2(t) \big] \frac{\hat{z}}{c}. 
\end{eqnarray}
It will be useful for a later discussion to have an explicit equation for the square of the nondipole effective momentum $\vbpi(\vbk,t)$, i.e.~
\begin{eqnarray}
\vbpi^2(\vbk,t) &=& k^2 + \vbA_0(t) \cdot [2 \vbk + \vbA_0(t) ]   \Big[1+ \frac{\vbk \cdot \hat{z}}{c} \Big]  + O(1/c^2) \label{Pi22}. 
\end{eqnarray}
The approach presented in this section allows the dipolar scheme to be easily recovered by simply taking the limit $1/c \rightarrow 0$. 
In fact, in such condition  Eq.~(\ref{volkovNONdipo}) goes to  the standard dipole Gordon-Volkov function \cite{Joachain12}. 

It is worth highlighting that higher-order corrections to the Hamiltonian are comparable to terms of order $c^{-2}$ derived from a purely relativistic Dirac and Klein-Gordon equations; see \cite{Forre2022,Forre2020,Forre2019}. 
Thus, the time-dependent Schr\"odinger equation of nonrelativistic quantum mechanics must be revised if terms of $O(1/c^2)$ were taken into account, which is out of the scope of the present article.
Here we want to mention that a semirelativistic formulation within the so-called propagation gauge \cite{Forre2016} was successfully derived \cite{Lindblom2018,Forre2019,Forre2020} in the transition from the dipole laser-matter interactions to a fully relativistic treatment required when the ponderomotive potential energy $U_p$ is comparable to the rest mass of the electron (see Fig.~\ref{fig1:tipoReiss}).

%--------------------------------------------------------
\subsection{Laser-assisted photoelectron emission (LAPE)}
%--------------------------------------------------------

\begin{figure}
   \centering
   \includegraphics[width=0.9\linewidth]{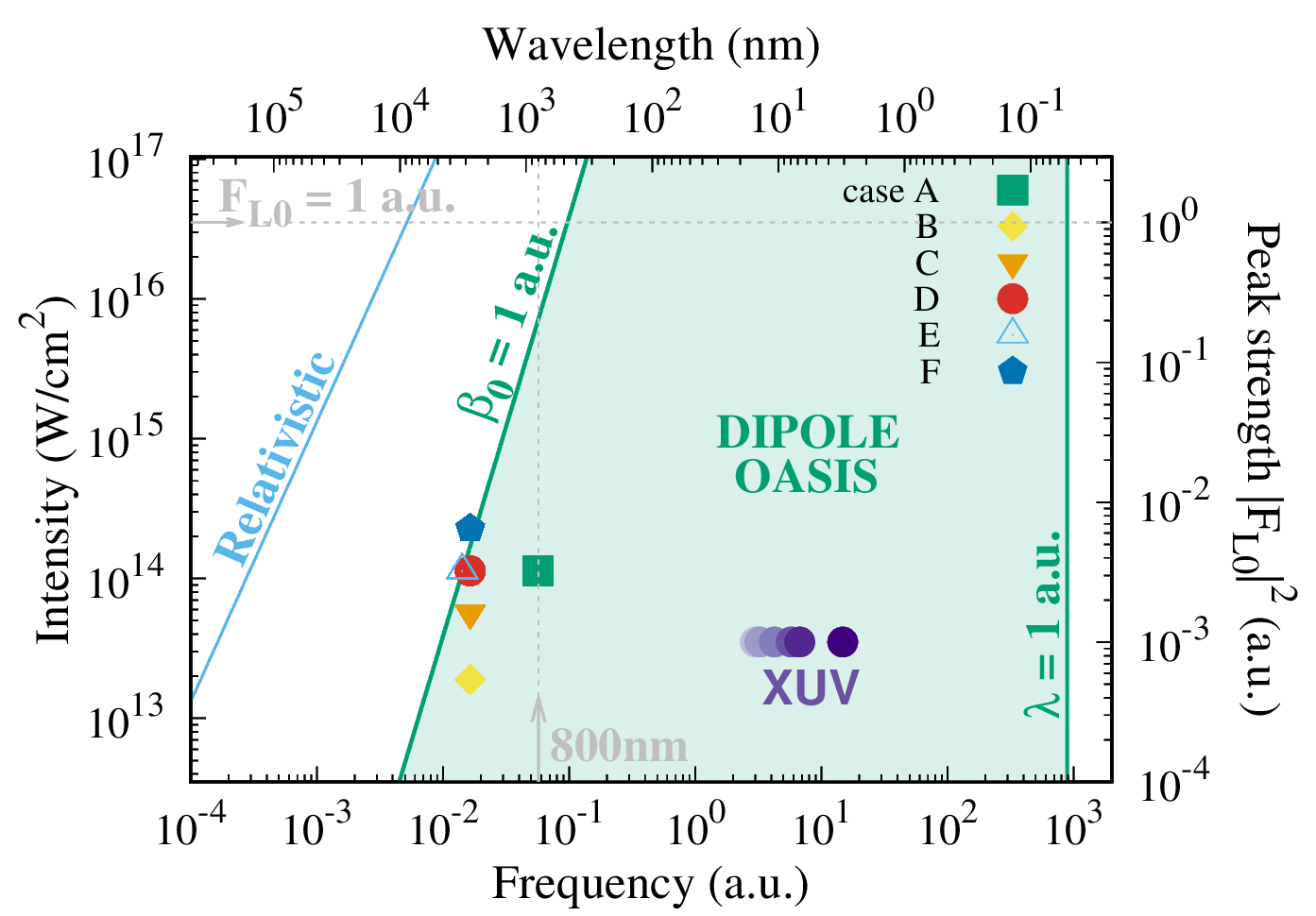}
   \caption{(Color online) Intensity-frequency map indicating the threshold curves for the validity of the dipole approximation (green region), the nondipole regime (above $\beta_0=1$ line), and the relativistic regime (above the light blue line defined as $U_p=mc^2$). 
   The points on the plane represent the parameter values used in the present calculations (see Table \ref{T1}).
   \label{fig1:tipoReiss}}
\end{figure}

We consider ionization of an atomic system by the combination of an extreme ultraviolet laser pulse assisted by an intense infrared or near infrared laser field.
In the single-active-electron (SAE) approximation, the time-dependent Schr\"odinger equation (TDSE) reads
\begin{equation}
i\frac{\partial }{\partial t}  \psi (t)  =
\Big[ H_0  + H_\mathrm{int}(t)  \Big]
\psi (t), 
\label{TDSE}
\end{equation}
where $H_0=-\nabla^2/2+V(r)$ is the time-independent atomic Hamiltonian. 
The first term represents the electron's kinetic energy, while the second term describes the electron-core Coulomb interaction.
In Eq.~(\ref{TDSE}), $H_\mathrm{int} = \vbE_X(t)\cdot \vbr + H_\mathrm{eEM}^L$ describes the interaction of the atom with both time-dependent XUV and IR  electric fields in the length gauge. 
We assume the XUV pulse to be sufficiently weak and of short wavelength, yet large enough compared to the atomic radius (see Fig.~\ref{fig1:tipoReiss} and its discussion), such that XUV ionization can be treated within the dipole approximation, with nondipole effects arising only from the subsequent action of the NIR laser. 
In Fig.~\ref{fig1:tipoReiss} (following the analysis by Reiss \cite{Reiss14}), we show the regions where the dipole approximation is well justified, along with the points representing the sets of parameters selected in this work. 
As observed, the XUV pulse considered here is clearly deep in the dipole regime.

%------------FIG:1----------------------------------------
%---------------------------------------------------

The electron, initially bound in an atomic state $\phi_{i}$, is emitted to a final continuum state $\phi_{f}$, with momentum $\vbk$ and energy $E=k^2/2$. 
The photoelectron momentum distribution (PMD) can be calculated as
\begin{equation}
\frac{\rmd P}{\rmd \vbk}= |T_{\infi}|^2 ,
\label{prob1}
\end{equation}
where $T_{\infi}$ is the $T$-matrix transition amplitude element corresponding to the transition $\phi_{i}\rightarrow\phi_{f}$. 
Within the time-dependent distorted wave theory, $T_{\infi}$ in the prior form can be written as
\begin{equation}
T_{\infi}= -i\int_{-\infty}^{+\infty}\rmd t \,\langle\chi_{f}^{-}(\vbr,t)|H_\mathrm{int}(\vbr,t)|\phi_{i}(\vbr,t)\rangle,
\label{Tif}
\end{equation}
where $\phi_{i}(\vbr,t)=\varphi_{i}(\vbr)\,e^{i I_{p} t}$ is the initial atomic state with ionization potential $I_{p}$, and $\chi_{f}^{-}(\vbr,t)$ denotes the distorted final state.
\Eref{Tif} is exact as far as the final channel, $\chi_{f}^{-}(\vbr,t)$, is the exact solution of Eq.~\eref{TDSE}.
However, various levels of approximation have been applied to solve Eq.~(\ref{Tif}). 
The most widely known is the strong field approximation (SFA), which neglects the Coulomb distortion caused by the interaction between the ejected electron and the residual ion in the final channel. 
Additionally, it disregards the influence of the laser field on the initial ground state~\cite{maciej1994,symphony}. 

In the present work, we approximate the distorted final state using the nondipole Gordon-Volkov wave function Eq.~\eref{volkovNONdipo}, where the space- and time-dependent IR laser EM field is defined in Eqs.~\eref{EM2}. 
Since the wave propagates in the $\hat{z}$ direction, the polarization remains in the $\hat{x}-\hat{y}$ plane.
Following Eq.~\eref{Anodip}, a right circularly polarized vector potential and electric fields can be written as
 \begin{eqnarray}
 \vbA_0(t)&=&\frac{F_{L0}}{\omega{\scriptstyle \sqrt{2}}} 
 [~ \hat{x } ~ \sin{(\omega t)} +\hat{y} ~ \cos(\omega t) ] \label{Avector} \quad \textrm{and} \\
 \vbE_0( t)&=&\frac{F_{L0}}{{\scriptstyle \sqrt{2}}} ~ [ -\hat{x }~  \cos{(\omega t)} +\hat{y } ~ \sin(\omega t) ],
\end{eqnarray}
where $F_{L0}$ is the electric field peak amplitude. 
The linearly polarized IR case was already considered in a previous work \cite{DellaPicca2023}.
For the XUV we consider a linearly polarized pulse of duration $\tau_X$, modeled as
\begin{equation}
    \vbE_{X}(t)= \left\{\begin{array}{ll}
        -\hat{\varepsilon}_X F_{X0}(t)\cos(\omega_{X}t) & \text{for }  \vert t-t_0 \vert\le \tau_X/2\\
        0  & \text{otherwise}
        \end{array}\right.
        \label{envelope}
\end{equation}
where $\hat{\varepsilon}_X$ and $\omega _{X}$ are the polarization vector and the carrier frequency, respectively. 
In Eq.~(\ref{envelope}), $F_{X0}(t)$ is an envelope function, nonzero only during the temporal interval $(t_{0}-\tau_X/2,t_{0}+\tau_{X}/2)$. 
Eventually, it can be approximated by its maximum amplitude, i.e.,~$F_{X0}(t)\approx F_{X0}$, typically reached at time $t_0$, the midpoint of the pulse indicating the XUV time delay with respect to the IR field.

The total vector potential is the sum of the IR and XUV ones, with the IR field dominating in terms of the amplitude-to-frequency ratio. 
Moreover, it has been shown in Ref.~\cite{DellaPicca13} that the so-called ``\textit{DipA}" approximation accurately describes the XUV single-photon absorption. 
This approximation is obtained by simply setting $\vbA_{XUV}\simeq 0$ in the Volkov phase. 
With this in mind, we set $\vbA_{tot} = \vbA_{IR} + \vbA_{XUV} \simeq \vbA_0$ as the total vector potential used in $\vbpi(\vbk,t)$ and the Volkov phase (see Eqs.~\eref{volkovNONdipo} and \eref{Pi22}).  

The transition amplitude can be split into two contributions: $T_{\infi}^{IR}$, the ionization by the IR laser, and $T_{\infi}^{XUV}$, the ionization by the XUV laser, as  $H_\mathrm{int} = \vbE_X(t)\cdot \vbr + H_\mathrm{eEM}^L$ consists of two terms (IR and XUV) contributions, i.e.,
\begin{equation}
  T_{\infi} = T_{\infi}^{XUV} + T_{\infi}^{IR}.  
\end{equation}
An interesting aspect of LAPE ionization processes is that, with an appropriate choice of IR and XUV laser parameters, the energy domains of XUV and IR induced ionization can be well separated, i.e.,  $|T_{\infi}^{XUV}|^2 + |T_{\infi}^{IR}|^2$. 
While $|T_{\infi}^{IR}|^2$ extends from the energy threshold up to $2U_p$\footnote{Except for linearly polarized dipole IR lasers, for which the $10U_p$ limit can be reached due to the HATI process~\cite{MiloJPBReview2006}.}
~\cite{MiloJPBReview2006}, the XUV term is centered at $\omega_X$ for one-photon absorption. 
Thus, by selecting sufficiently high XUV frequencies, the two domains do not overlap. 
Hereinafter, we focus exclusively on $T_{\infi}^{XUV}$, leaving the analysis of the IR term. 
Thus, in order to study $T_{\infi}^{XUV}$, we can assume $H_\mathrm{int} \simeq  \vbE_X(t) \cdot \vbr$.
Considering only XUV photon absorption\footnote{Disregarding the emission of one XUV photon we can consider $\cos(\omega_X t) \simeq \exp{(-i \omega_X t)}/2$.},
the matrix element in Eq. \eref{Tif} can be written as
\begin{equation}
T_{\infi} \simeq T_{\infi}^{XUV}   =  \int_{t_0-\frac{\tau_X}{2}}^{t_0+\frac{\tau_X}{2}}  \ell(\vbk,t) \, \exp{[iS(\vbk,t)]} \,\,\rmd t,  \label{Tifg}
\end{equation}
where
\begin{eqnarray}
\ell(\vbk,t) & = &  -  \frac{i}{2}  F_{X0}(t) \, \hat{\varepsilon}_X \cdot  \vbd \big[ \vbpi (\vbk,t) \big] \, \label{elle},
\end{eqnarray}
and $S(\vbk,t)$ is the generalized action
\begin{eqnarray}
S(\vbk,t) &=& (I_{p} -\omega_{X}) t + \frac{1}{2} \int^t \rmd t^\prime \vbpi^2(\vbk,t^\prime).
\label{S1}
\end{eqnarray}
In Eq.~\eref{elle} we have introduced the dipole element $\vbd [\vbv] = \frac{1}{(2\pi)^{\frac{3}{2} } } \langle e^{i \vbv \cdot \vbr} \vert \vbr \vert \varphi_{i}(\vbr) \rangle$ between an atomic state $\varphi_i(\vbr)$ and an electronic plane wave with velocity $\vbv$.

Keeping in mind Eqs.~(\ref{Pi22}) and (\ref{Avector}), and neglecting terms of order $1/c^2$, the generalized action can be written as
\begin{eqnarray}
S(\vbk,t) &=& a t + b \cos (\omega t) + b^\prime \sin (\omega t)  \label{S},
\end{eqnarray}
where
\begin{subequations}
\begin{eqnarray}\label{abc} 
a &=&   \frac{k^{2}}{2}+I_{p} - \omega_{X} +U_p \Big[1+\frac{\vbk\cdot\hat{z}}{c}  \Big], \\
b &=&  \frac{-F_{L0}}{\omega ^{2} {\scriptstyle \sqrt{2}}} (\vbk \cdot \hat{x}) \Big[ 1+\frac{\vbk \cdot \hat{z}}{c}  \Big]\quad \textrm{and} \\
b^\prime &=&  \frac{F_{L0}}{\omega ^{2} {\scriptstyle \sqrt{2}}} (\vbk \cdot \hat{y}) \Big[ 1+\frac{\vbk \cdot \hat{z}}{c}  \Big].   
\end{eqnarray}
\end{subequations}
Here, the parameter $U_p=(F_{L0} /2 \omega)^2$ represents the ponderomotive energy. 

The PMD can be calculated using Eq. (\ref{prob1}) and Eq.~\eref{Tifg}. 
However, it is insightful to examine the periodicity properties of the transition-matrix amplitude, as previously discussed in \cite{DellaPicca2020}. 
It is straightforward to observe that, since the vector potential $\vbA_0(t)$ oscillates with a period $T=2\pi/\omega$, the action $S(\vbk,t)$, the effective momentum $\vbpi$, and the dipole element $\vbd$ all exhibit $T-$periodicity of the form
\begin{eqnarray}
    \vbd[\vbpi(\vbk,t+jT)] &=&     \vbd[\vbpi(\vbk,t)]  \label{propd} \\
    S(\vbk,t+jT) &=& S(\vbk,t) + a j T \label{propS} ,
  \end{eqnarray}
where $j$ is an arbitrary integer. 
Note that the present Eqs.~\eref{propd} and \eref{propS} are equivalent to Eqs.~(9) and (12) in Ref. \cite{DellaPicca2020} where we derived analytical expressions for PMD within the dipole approximation, corresponding to typical regimes such as streaking, sidebands, and pulse trains.
With an appropriate choice of XUV profile envelope, the analytical expressions previously derived in Ref. \cite{DellaPicca2020} remain valid in the present nondipole approach.
%In particular, Eq.~(33) of \cite{DellaPicca2020}, which corresponds to the attosecond pulse train (APT) regime.

%--------------------------------------------------
\subsection{Attosecond Pulse Train}    
%-------------------------------------------------

An attosecond pulse train (APT) can be interpreted either as a specific sum of harmonics in the frequency domain (see, for example, \cite{JimenezGalan2013}) or as a field composed of a series of small pulses that repeats periodically in the time domain (see, for example, Fig.~1c of \cite{DellaPicca2020}). 
Let us consider the case of an APT with the same $T-$periodicity of the IR laser, i.e., a short XUV pulse that repeats every optical cycle of the IR field. 
Each $j$th individual pulse has the shape given in Eq.~\eref{envelope}, centered at $t_{0,j} =t_0 + (j-1) T$, with $j$ ranging from $1$ to $N$, the total number of pulses. 
Then, $\ell$ also fulfills
\begin{equation}
    \ell(t+jT)= \ell(t)  \label{propl}
\end{equation}
and analytical expressions derived in Ref. \cite{DellaPicca2020} are applicable in this context.
In particular, according to Eq.~(33) of \cite{DellaPicca2020}, the emission probability for the pulse train can be expressed as the product of two interference contributions: (i) the intrapulse interference factor corresponding to the emission probability of a single isolated pulse, i.e. the streaking pattern,
and (ii) the interpulse interference accounting for the coherent interference between emissions from different pulses of the train \cite{Gramajo18,DellaPicca2020}. 
Then,
\begin{equation}
 |T_{\infi}|^2 =   \underbrace{|T_{\infi}^{\mathrm{st}}|^2}_{\mathrm{intrapulse}} \, 
 \underbrace{
 \left[\frac{\sin{( a T N /2)}}{ \sin{(a T/2) }}\right]^2
 }_{\mathrm{interpulse}} . \label{eqtren}
\end{equation}
If the repetition rate of the pulse train matches the IR cycle, as in the present case, the interpulse interference factor resembles the intercycle interference factor, proper of the sideband regime \cite{DellaPicca2020,DellaPicca2023}.

\begin{figure}
    \centering
    \includegraphics[width=0.75\linewidth]{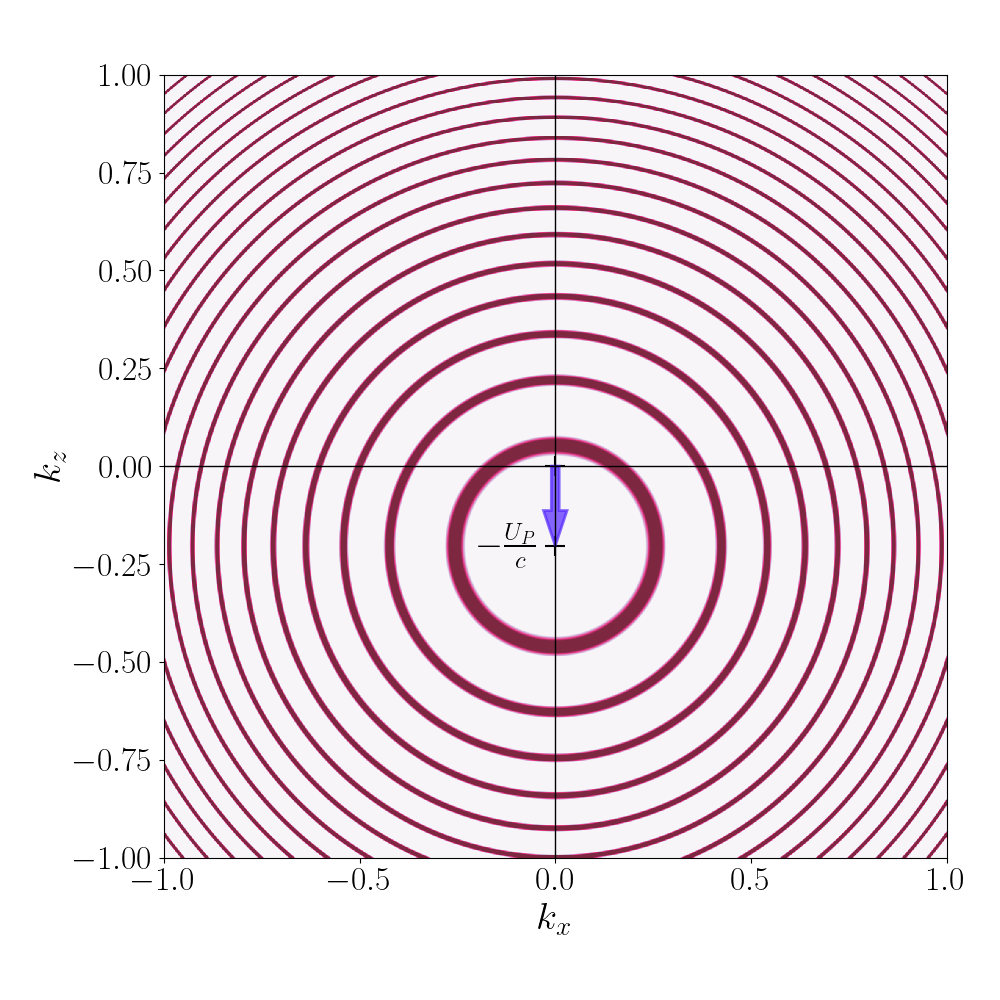}
    \caption{(Color online) Sideband rings in momentum space. 
    For this figure we use $c=13.7$ a.u. to enhance the shift in the center of symmetry of the interpulse interference rings given by Eq. (\ref{eqtren}). }
    \label{fig2:anillosSB}
\end{figure}

Let us first analyze the inter-pulse/cycle factor. 
The zeros of the denominator, i.e., the energy values satisfying $a T/2 =n\pi$, are removable singularities, as the numerator also cancels out, resulting in maxima at these points. 
These maxima correspond to sideband peaks in the photoelectron spectrum and exhibit angle dependence or inclination, which increases with energy \cite{DellaPicca2023}. 
Even more, using Eq. (\ref{abc}) and rewriting the condition $a T/2 =n\pi$ in momentum space, we obtain
\begin{equation}
   k_{x}^2 + k_{y}^2 
   + \Big(k_{z} +\frac{U_p}{c}\Big)^2 =  2 \big(  n \omega_L + \omega_X -I_p-U_p \big)+ O(1/c^2).\label{spherequations} 
\end{equation}
This equation can be interpreted as a sphere in momentum space for each $n$, with radius $\sqrt{2(n\omega_L +\omega_X-I_p-U_p)}$ and shifted an amount $U_p/c$ in the $-\hat{z}$ direction (see Fig.~\ref{fig2:anillosSB}).
One would expect a shift in the same direction as the IR laser propagation, 
due to the photon momentum transfer, and not opposite to it. 
Thus, each sphere given by Eq. (\ref{spherequations}) must be anisotropically modulated so that the total momentum is shifted in the propagation direction, as expected.
This observation may be consistent with the findings of F\o{}rre \cite{Forre2022} in strong field ionization by infrared light. 
He has found that the quadratic vector potential operator (whose average is essentially the ponderomotive energy $U_p$ \cite{DellaPicca2016}) is responsible for a negative induced nondipole shift of the photoelectron momentum opposite to the light propagation direction, whereas another linear term is generally responsible for a corresponding positive shift. 

%----------------FIG 2-----------------------------------
%---------------------------------------------------

The intrapulse factor corresponds to a single isolated XUV pulse and, in this sense, resembles a streak-camera scenario, i.e., a streaking PMD, and hence the ``st" superscript.  
In this case, we carry out the time integration into the transition matrix $T_{\infi}^{\mathrm{st}}$ in \Eref{Tifg} from the beginning of the isolated XUV pulse, $t_0- \tau_X/2$, to its end, $t_0+\tau_X/2$. 
As mentioned earlier, we eventually assume that the envelope of the XUV field reaches its maximum amplitude, $F_{X0}$, at the delay time $t_0$. 
Given the short time span, it can also be assumed that ionization occurs instantaneously, i.e., at the specific time $t_0$.

With the above analysis in mind, we can proceed analytically using the semiclassical
model (SCM).
The SCM involves evaluating the time integral of Eq.~\eref{Tifg} using the saddle-point approximation (SPA), where the primary contribution comes from times $t_s$ at which the action is stationary, i.e., $dS/dt =0$. 
This stationary time $t_s$ can be attributed to the ionization time $t_0$, which means that in the streaking regime, we set $t_s = t_0$.
Then,  the integral \eref{Tifg} can be approximated by evaluating it at a single  time, i.e.,
%
%\begin{equation}
%T_{\infi}^{\mathrm{st}}  =  \int_{t_s-\tau_X/2}^{t_s+\tau_X/2} \ell(t) 
%\, \rme^{iS(t)} \,\,\rmd t  
%\, \simeq  \,  \sum_{t_s} \ell(t_s) ~e^{i(S(t_s)+ \frac{\pi}{4}\ddot{S}(t_s))} ~ \sqrt{\frac{2 \pi}{|\ddot{S}(t_s)|}}.
%\end{equation}
%
\begin{equation}
T_{\infi}^{\mathrm{st}}  =  \int_{t_s-\tau_X/2}^{t_s+\tau_X/2} \ell(t) 
\, \rme^{iS(t)} \,\,\rmd t  
\, \simeq  \,   \ell(t_0) ~e^{i(S(t_0)+ \frac{\pi}{4}\ddot{S}(t_0))} ~ \sqrt{\frac{2 \pi}{|\ddot{S}(t_0)|}}.
\end{equation}
where, from Eq.~\eref{S1} we have $\dot{S}(t) =  - v_0^2 /2 + \boldsymbol{\Pi}^2(\vbk,t) /2$ and  $\ddot{S}(t) = \boldsymbol{\Pi}(\vbk,t)  \cdot \dot{\boldsymbol{\Pi}}(\vbk,t)$.
The streaking PMD becomes then proportional to $|T_{\infi}^{\mathrm{st}}|^2 \simeq  2\pi |\ell(t_0)|^2   |\ddot{S}(t_0)|^{-1}$ under the saddle condition  $dS/dt=0$ at time $t_0$, which yields
\begin{equation}
 \boldsymbol{\Pi}^2(\vbk,t_0) = v_0^2,  \label{dsdt0}
\end{equation}
where $v_0^2/2=\omega_X-I_p $ corresponds to the mean energy of photoelectrons ionized by the XUV pulse in absence of the NIR laser field.
Unlike the SCM in the sideband regime, where an entire region in momentum space is accessible for multiple values of $t_s$ (see \cite{DellaPicca2023}), in the streaking regime, only a few momentum values are permitted by Eq.~\eref{dsdt0} for a single $t_0$:  $\vbk(t_0)$. 
Within the dipole approximation, these values define a spherical surface in momentum space with a radius $v_0$ and centered on $-\vbA_0(t_0)$. 
However, the nondipole surface $\vbk(t_0)$ is not exactly spherical but slightly deformed due to term of order $1/c$ in Eq.~\eref{Pi22}.

%--------------FIG 3-----------------------------------
\begin{figure}
  \centering
  \includegraphics[width=0.85\linewidth]{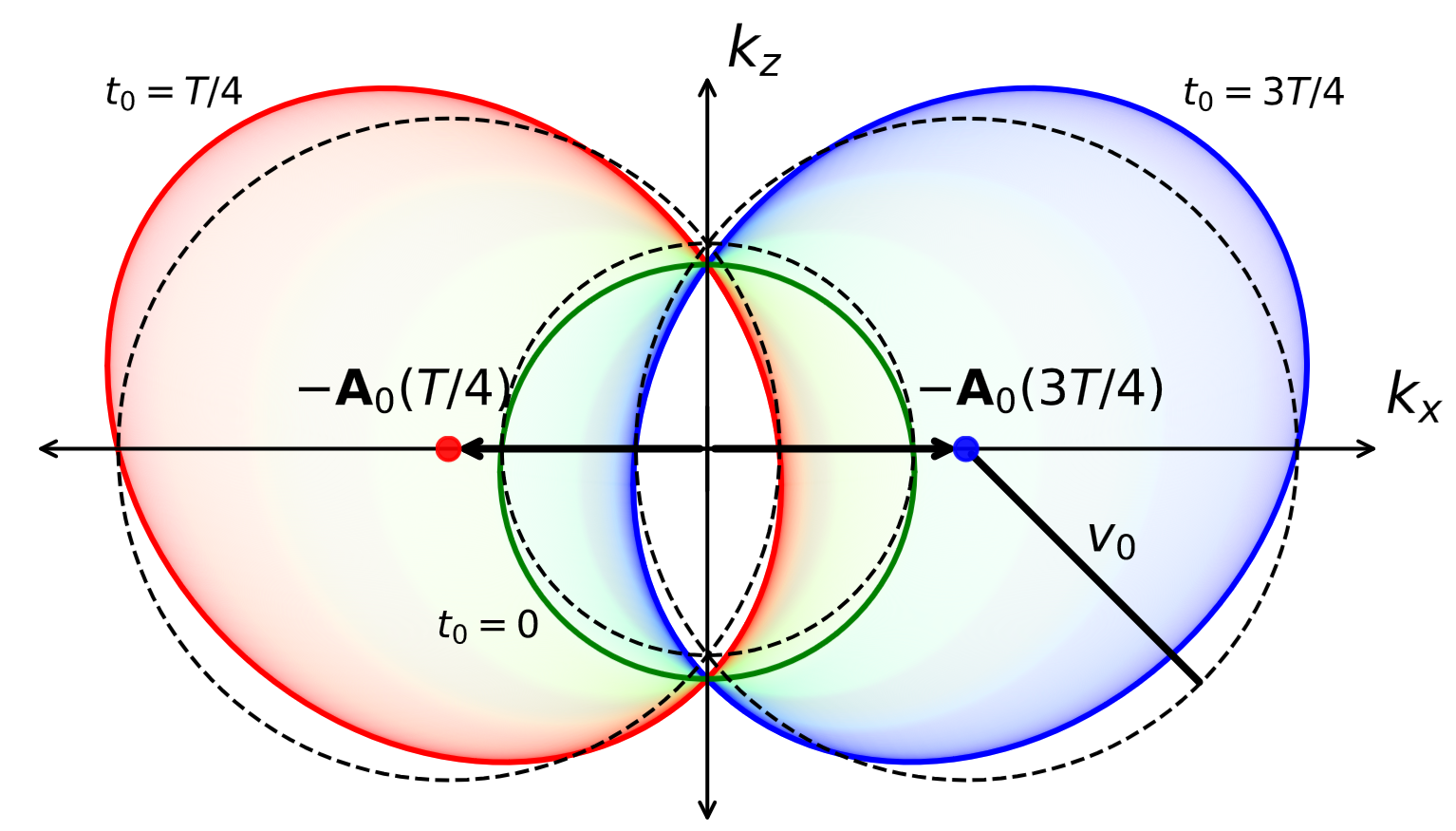}
  \caption{ (Color online) Intersection between \( \vbk(t_0) \) surfaces and the \( k_y=0 \) plane for several values of \( t_0 \), where \( c=13.7 \, \mathrm{a.u.} \) was considered to highlight the non-dipolar asymmetries. 
  Center (green): \( t_0=0 , T/2 \) and \( T \). 
  Left (red) and right (blue) correspond to \( t_0=T/4 \) and \( 3T/4 \), respectively. 
  Dashed lines: SCM within the dipole approximation.}
  \label{fig3:SCM}
\end{figure}
%------------------------------------------------

In Fig.~\ref{fig3:SCM}, we show the solutions of Eq.~(\ref{dsdt0}) for different values of $t_0$, projected onto the $k_y = 0$ plane and in order to emphasize the differences from the dipole approximation, the speed of light has been artificially shortened to $c=13.7$, tenth of its real value. 
As $t_0$ varies from $0$ to $T$, the vector potential traces a full circumference in the $z=0$ plane [see Eq.~\eref{Avector}], causing the values of $\vbk$, which are solutions of Eq.~\eref{dsdt0}, to evolve over time. 
The solutions depicted in Fig.~\ref{fig3:SCM} enable us to interpret the structure of the classical solution in three-dimensional momentum space. 
The resulting structure corresponds to an ovoid surface whose center rotates clockwise around the $\hat{z}$-axis.
At $t_0 = 0$, $-\vbA_0(t_0)$ points opposite to the $\hat{y}$-direction, and the $k_x$ and $k_z$ components form the central green circular contour, subtly displaced downward along the $k_z$-axis, opposite to the propagation direction of the IR laser. 
As $t_0$ increases from $0$ to $T/4$, the region shifts leftward, enlarging the enclosed area and introducing a forward-backward asymmetry in the $k_z$ direction of the electron emission. 
Compared to the dipole approximation, the momentum magnitude is intensified in the second quadrant and reduced in the third one, see the left (red) curve. 
For $T/4 \leq t_0 \leq T/2$, the process reverses, returning to the central green circular region. 
A similar process occurs to the right as $t_0$ progresses, reaching the right (blue) contour at $t_0 = 3T/4$ before returning to the center for $t_0=T$.
It is only at $t_0 = T/4$ and $t_0 = 3T/4$ that $-\vbA_0(t_0)$ lies in the plane of the figure. 
For these particular times and $t_0=0$, the solutions in the dipole approximation correspond to spheres (circumferences in the $\hat{x}-\hat{y}$ plane), shown as black dashed lines in the figure.

%------------------------------------------------
\section{Results} \label{Results}
%-----------------------------------------------

We present the ionization spectra for a pulse train consisting of only $N=2$ XUV pulses. 
In order to perform the calculation of $T_{\infi}^{\mathrm{st}}$ according to  \Eref{Tifg}, each isolated short XUV pulse, will be assumed to have a sin$^2$-shaped envelope of duration of $\tau_X = T/6$, i.e.,
\begin{equation}
F_{X0}(t) = F_{X0} ~\sin^2\left[\frac{\pi (t-t_0)}{\tau_X}\right].
\label{envelopeX}
\end{equation}
We select a series of parameter sets such that the calculations progressively deviate from the dipole regime, with increasing $\beta_0$ as shown in Table \ref{T1}. 
The XUV polarization is assumed to be aligned with the direction of IR propagation, $\hat{\varepsilon}_X = \hat{z}$.

%------------TABLE 1-----------------------------
\begin{table}
\caption{\label{T1} Laser parameters for each study case. In all the cases $F_{X0}=0.01$, $\hat{\varepsilon}_X=\hat{z}$ } 
\footnotesize
\begin{tabular}{@{}lllllll}
\br
CASE&$F_{L0}$ (I in W/cm$^2$)& $\omega $ ($\lambda$ in nm) &$U_p$&$\omega_X$& $v_0$ &$\beta_0=U_p /2\omega c $\\
\mr
A & $0.0569$ (1.14 x 10$^{14}$ W/cm$^2$) & $0.0569$ (800.8 nm) &$1/4$   &  $3$ & 2.24 & $2.5/c$ \\
B & $0.0569/\sqrt{6}$ (1.89 x 10$^{13}$ W/cm$^2$) & $0.0569/\sqrt{12}$ (2773.9 nm)  &$1/2$   &  $3.25$ & 2.34 & $15.2/c$\\
C  & $0.0569/\sqrt{2}$ (5.68 x 10$^{13}$ W/cm$^2$) & $0.0569/\sqrt{12}$   &$3/2$ & 4.25 & 2.74 & $45.7/c$\\
D & $0.0569$  & $0.0569/\sqrt{12}$   &$3$ & 5.75 &3.24  & $91.3/c = 0.67$\\
E & $0.0569$ & $0.0569/\sqrt{16}$ (3203 nm) &$4$ & 6.75 & 3.54 & $140.6/c=1.03$\\
F & $0.0569 \times 2$ (4.55 x 10$^{14}$ W/cm$^2$) & $0.0569/\sqrt{12}$ & 12 &$14.75$ & 5.34 &  $365.3/c=2.67$\\
\br
\end{tabular}
\end{table}
%-----------------------------------------
\normalsize

Although incorporating complex atomic states would not be difficult, we have simplified the calculation by considering a $1s$ hydrogenic state, for which
\begin{eqnarray}
\vbd_{1s} (\vbv) &=& -\frac{i}{\pi}\, 2^{7/2}\alpha^{5/2}\, 
\frac{ \vbv}{ (v^2+\alpha^2)^3}  \label{1s}
\end{eqnarray}
where $\alpha = \sqrt{2I_p} =1 $ since $I_p =13.6058$eV $= 0.5$ a.u. for an H($1s$) atom. 

\begin{figure}
   \centering
   \includegraphics[width=0.4\linewidth]{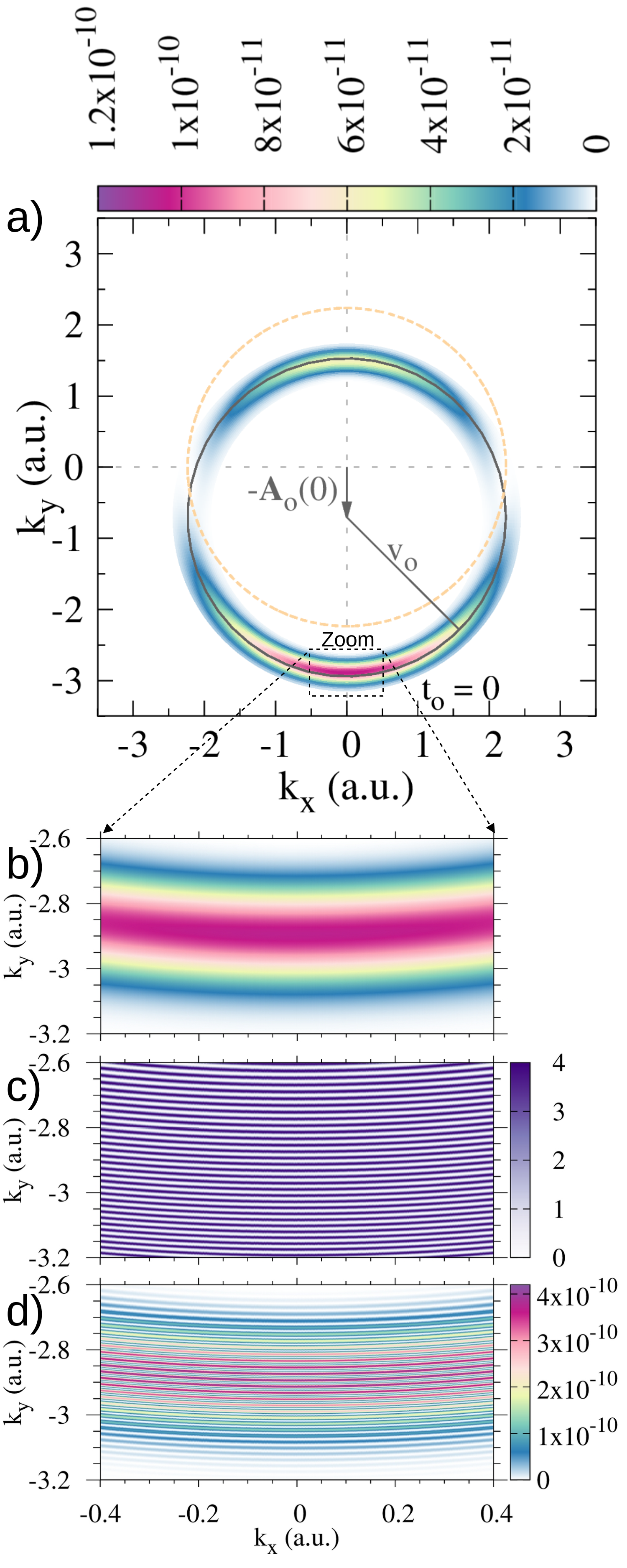}
   \caption{(Color online) a) Streaking PMD for H($1s$) in case A (see Table \ref{T1}) in the $k_z=0$ plane for $t_0=0$. 
   The SCM prediction is represented by the black line, and the Cooper-like minima are indicated by the orange dashed line. 
   Refer to the text for further details. 
   a) Zoomed-in view of the intrapulse factor, b) the interpulse contribution and c) the product of inter- and intrapulse factors.} 
   \label{fig4:kz0}
\end{figure}

\begin{figure}
   \centering
   \includegraphics[width=1\linewidth]{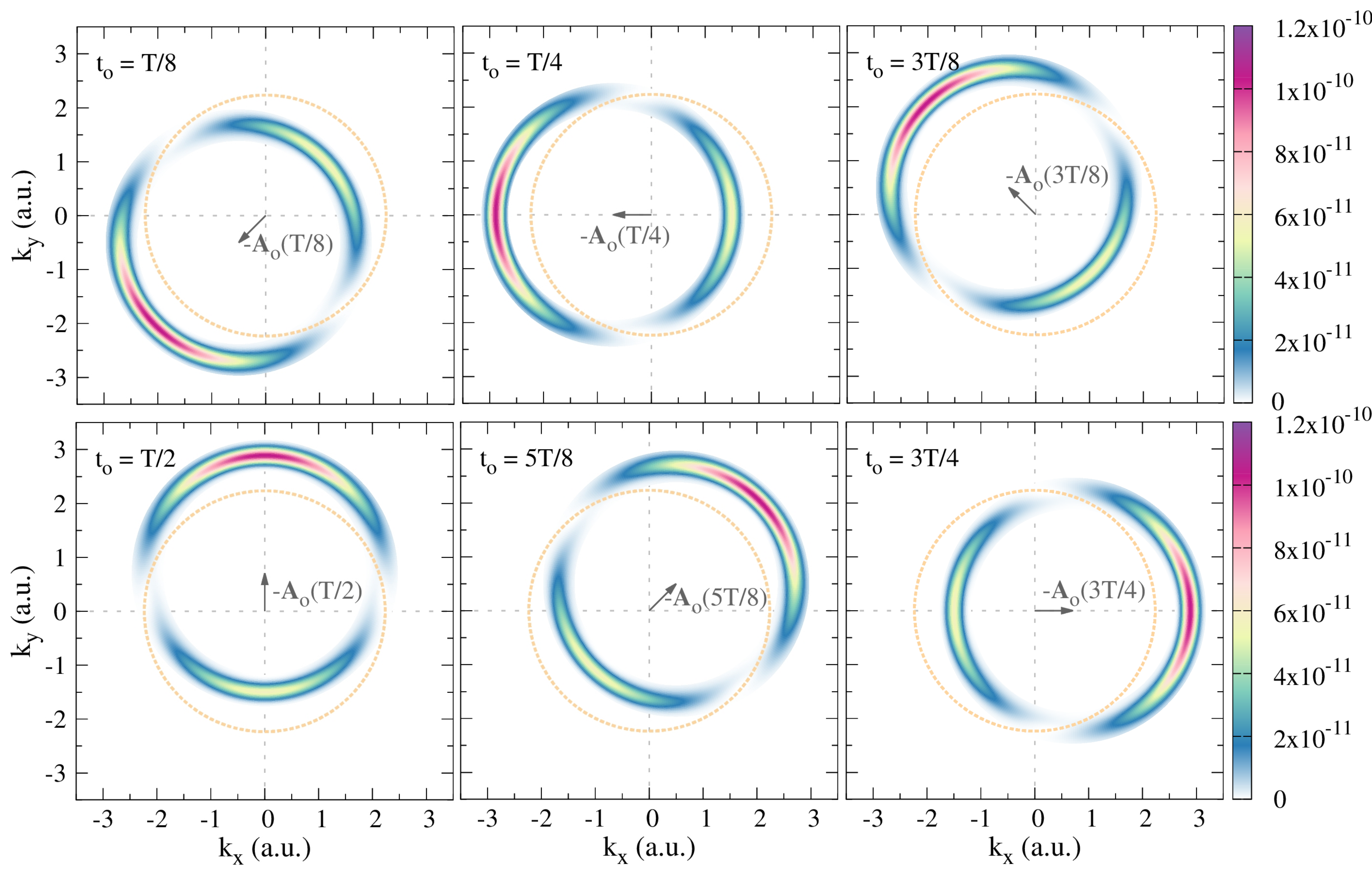}
   \caption{(Color online) Same as Fig.~\ref{fig4:kz0} a), but changing the ionization time $t_0$, indicated at the top left in each panel.}
   \label{fig5:kz0atAvstime}
\end{figure}

\begin{figure}
    \centering
    \includegraphics[width=0.97\linewidth]{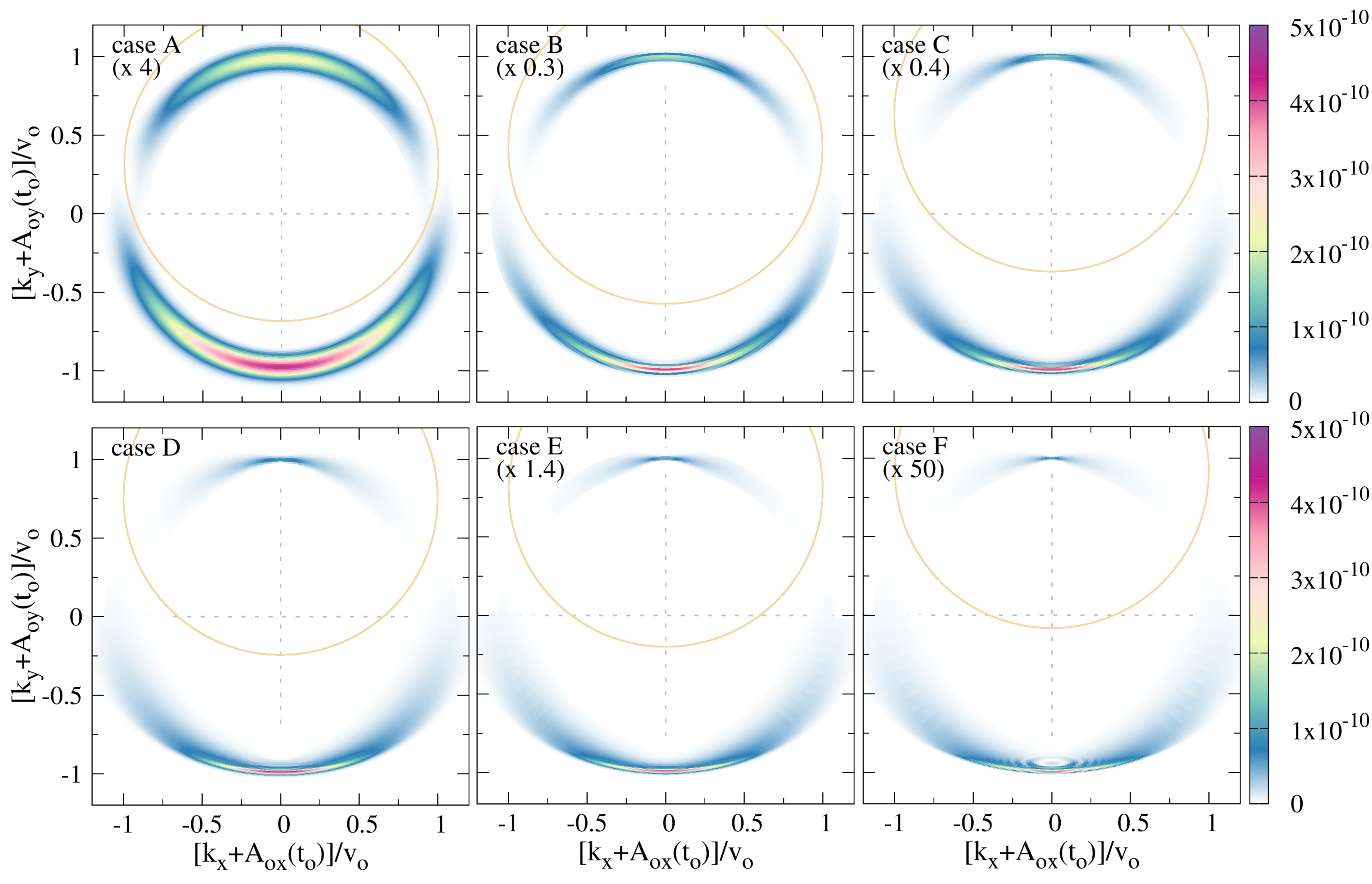}
    \caption{(Color online) Same as Fig.~\ref{fig4:kz0} a), but using the parameters labeled A to F in Table~\ref{T1}. 
    The momentum space is scaled with $v_0^{-1}$ and shifted by $-A_0(t_0=0)$, so the PMD emission approximately aligns with centered circumferences of radius 1.
    To maintain the same color bar, the streaking PMD are multiplied by 4, 0.3, 0.4, 1, 1.4 and 50 for cases A to F, respectively, as indicated in parentheses of each panel.    }
    \label{fig6:kz0.Morecases}
\end{figure}

\begin{figure}
\centering
\includegraphics[width=0.4\textwidth]{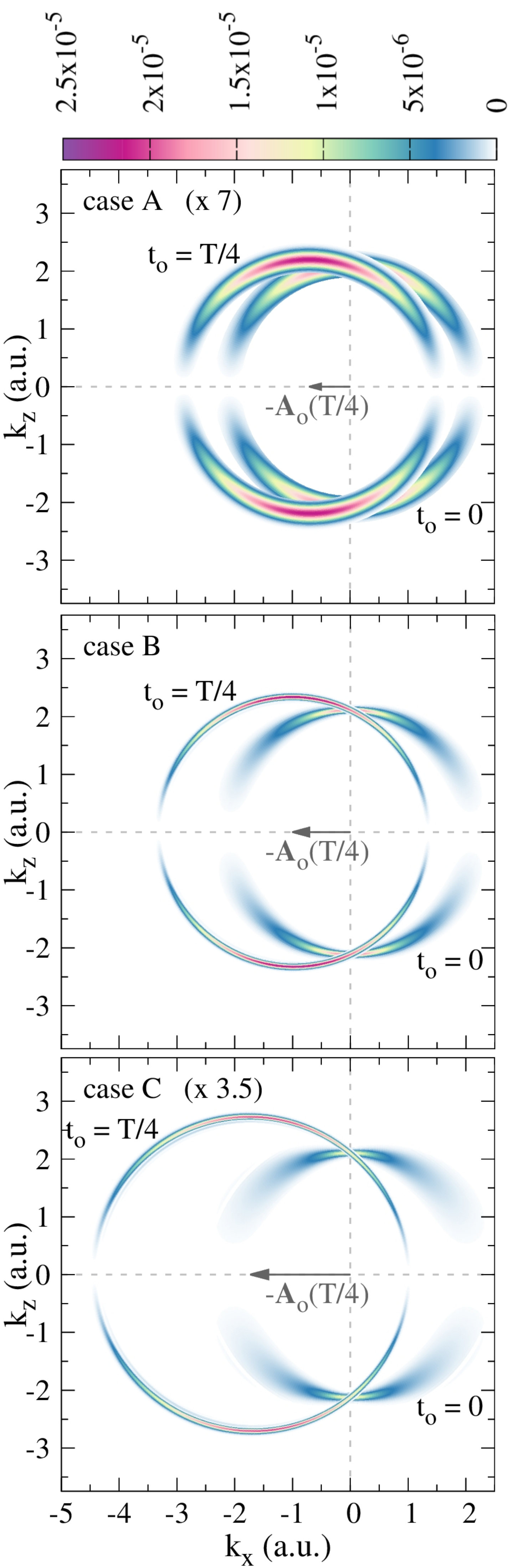}  
\caption{(Color online) Streaking PMD for cases A to C shown in the $(k_x,0,k_z)$ plane for emission times $t_0= 0$ (rings centered at the origin) and $T/4$ (rings centered at $-\vbA_0(T/4)$). 
Cases A and C are multiplied by $7$ and $3.5$ respectively.}
\label{fig7:ky0.MoreCases1}
\end{figure}

\begin{figure}
\centering
\includegraphics[width=0.33\textwidth]{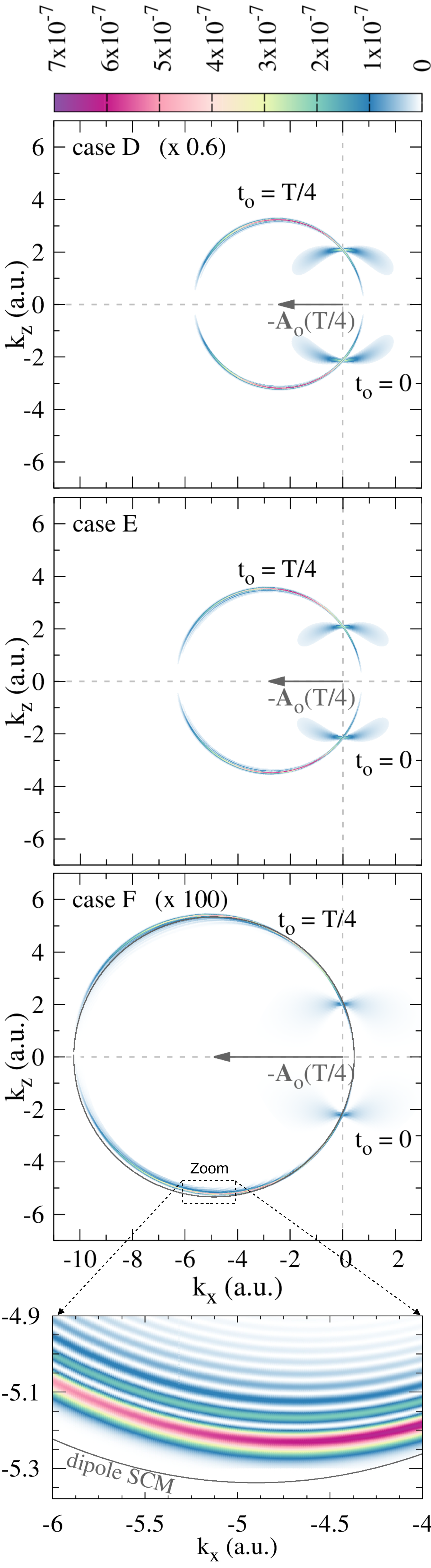} 
\caption{(Color online) Same as Fig.\ref{fig7:ky0.MoreCases1}, but using the parameters labeled D to F in Table~\ref{T1}. 
Cases D and F are multiplied by $0.6$ and $100$ respectively. 
The SCM prediction is represented by the black line.
Bottom: zoomed-in view of PMD for case F. } 
\label{fig8:ky0.MoreCases2}
\end{figure}

To compute the presented results, it was necessary to evaluate $\ell (\vbk, t)$ [Eq.~\eref{elle}], using Eq.~\eref{1s} for $\vbd_{1s}$, i.e.,
\begin{equation}
\ell (\vbk,t) = -\frac{F_{X0}(t)}{\pi} (2 \alpha)^{5/2} \frac{\hat{z} \cdot  \vbpi(\vbk,t)}{( \vbpi^2(\vbk,t)  + \alpha^2)^3 }. \label{Apen_ell}  
\end{equation}
In the following, we represent the PMD defined in Eq.~(\ref{prob1}) using the factorization of the streaking and interpulse interferences evidenced in Eq. (\ref{eqtren}).
According to this equation the streaking factor $|T_{\infi}^{\mathrm{st}}|^2$ modulates the (very fine spaced) ring-shaped sidebands arising from the interpulse interference.
Once it was understood how this modulation works (see Fig.~\ref{fig4:kz0}b, c and d and their discussion) we will concentrate in the PMD of the intrapulse factor for the rest of the figures.
To plot the PMD, which depends on the three momentum components, we choose to make cuts in two distinct planes: in Figs.~\ref{fig4:kz0}, \ref{fig5:kz0atAvstime}, and \ref{fig6:kz0.Morecases}, we fix $k_z = 0$, visualizing the $(k_x, k_y,0)$ plane. 
Contrariwise, in Figs. \ref{fig7:ky0.MoreCases1} and  \ref{fig8:ky0.MoreCases2}, we fix $k_y = 0$, displaying the $(k_x,0, k_z)$ plane, which corresponds to the same plane shown in the schematics of Figs~\ref{fig2:anillosSB} and \ref{fig3:SCM}.

%--------------FIG 4-------------------------
%---------------------------------------------------

In Fig.~\ref{fig4:kz0}, we present the $k_z=0$ PMD, corresponding to the case where the isolated XUV pulse is centered at $t_0 = 0$ associated with the parameter set labeled as A in Table \ref{T1}.
Note that the dipole limit is obtained by neglecting terms of the order $1/c$ or higher. 
According to Eqs.~(\ref{Apen_ell}) and (\ref{Pi22}), in the dipole limit one obtains $\ell \simeq \vbk \cdot \hat{z}$, since $A_0(t)$ stays perpendicular to the propagation direction, which implies that, emission in the direction perpendicular to the $\hat{z}$ axis is forbidden, as the transition matrix element becomes zero. 
However, emission does occur in that direction when nondipole contributions are considered. 
Consequently, the emission in the $k_z=0$ plane (as shown in Fig. \ref{fig4:kz0}) is inherently a nondipole effect \textit{per se}. 
In panels b) c) and d) of Fig.~\ref{fig4:kz0} we show a zoomed section of the PMD for the intrapulse factor, interpulse contribution and their product, respectively.
Since the small frequency value, the sidebands rings are very close to each other and thus in a general view, the structure of the streaking pattern survives: Fig. \ref{fig4:kz0}d recovers Fig. \ref{fig4:kz0}b. 
In what follows we analyze only the streaking PMD factor.  

Within the SCM, the momentum distribution is localized on a surface described by Eq. (\ref{dsdt0}) that, under the restriction $k_z=0$ turns out  $\left[\vbk +\vbA_0(t_0)\right]^2 = v_0^2$, which corresponds to a circumference of radius $v_{0}$ centered at $-\vbA_0(t_0)$. 
This is illustrated in Fig.~\ref{fig4:kz0}a, where the black line represents a circumference with a radius of $\sqrt{5}$, centered at $-\vbA_0(0)=-\frac{1}{\sqrt{2}} \hat{y}$. 
Over time, the center of the ring rotates clockwise, as shown in Fig.~\ref{fig5:kz0atAvstime}.
This Fig. corresponds to different polarization directions as a function of the delay time $t_0$, which is indicated in each panel. As $t_{0}$ increases, the instantaneous polarization of the IR pulse rotates and, thus, it does the center of the rings. 
It can be seen in all the panels that the emission is not symmetric as there is more intensity in the $\mathbf{k}$-region pointing to the $-\vbA_0$ direction. 
This can be considered as a nondipole angular streaking or attoclock process.
A very important result is that whereas the interpulse sidebands are shifted in the direction opposite to the IR propagation, the streaking pattern is shifted in the direction opposite to the instantaneous polarization direction of the IR. 
Since polarization and propagation directions are orthogonal in a propagating EM wave, both shifts are in orthogonal planes.  

%-----------FIG 5----------------------------------------
%---------------------------------------------------

From Figs.~\ref{fig4:kz0} and ~\ref{fig5:kz0atAvstime}, it is evident that all PMD spectra exhibit regions with minimal angular emission zones. 
This can also be interpreted within the SCM framework by incorporating Eqs. (\ref{dsdt0}) and (\ref{Apen_ell}):
\begin{equation}
\ell(t_0) =- \frac{F_{X0}(t_0)  (2 \alpha)^{5/2}}{ \pi (2\omega_X)^3 } \left[k_z + \frac{v_0^2 -k^2 }{2c(1+k_z/c)}\right] + O (1/c^2) .
\label{elle_ts}
\end{equation}
Under condition $k_z=0$ and dismissing terms of order $c^{-2}$, Eq. (\ref{elle_ts}) vanishes along the $E=v_0^2/2$ ring, represented by the dotted line in Figs. \ref{fig4:kz0} and \ref{fig5:kz0atAvstime}. 
This structure corresponds to a null region of the transition-matrix amplitude in momentum space.
In atomic photoionization, the zeros of the transition matrix are called \textit{Cooper minima} when they originate in the nodes of the initial wave function \cite{Cooper1962}. 
In both the present case and the photoionization of H$_2^+$ case (see \cite{DellaPicca2008,DellaPicca2009}), the initial state does not have any nodes, and for that we will refer to them as Cooper-\textit{like} minima.
It is worth highlighting that the factor $\ell(t_0)$ in Eq. (\ref{elle_ts}) is independent of the instantaneous polarization state of the IR laser, as we can observe in Fig.~\ref{fig5:kz0atAvstime}. 
This is the same reason why we have previously found the same condition $E=v_0^2/2$ for the Cooper-\textit{like} minimum in the case of linear polarization even in the sideband regime (see \cite{DellaPicca2023}). 
However, we expect it to be highly dependent on the initial orbital through the matrix element $\vbd[\vbpi(\vbk,t_0)]$, which is involved in $\ell(t_0)$. 
It would be interesting to investigate the Cooper-\textit{like} minimum conditions for different initial states in upcoming work.  

%-----------------------FIG 6----------------------------
%---------------------------------------------------

Figure \ref{fig6:kz0.Morecases} shows the streaking PMD for the different kinematical cases of Table \ref{T1} where the nondipole parameter $\beta_0$ increases from cases A to F.
The axes are shifted and scaled in the $(k_{x},k_{y},0)$ plane such that all the PMD are centered at the origin with radii equal to one. 
Thus, the rings maintain the same size for all panels changing their shapes, widening or narrowing in different zones, increasing the north-south asymmetry with increasing $\beta_0$. 
For higher values of $\beta_{0}$ the PMD emission concentrates in the direction pointed by the instantaneous polarization ($-\vbA_0(t_0)$) for $t_0=0$, direction in which it also narrows (in scaled units). 
As $\beta_0$ increases from case A to case F, the PMDs become narrower and a blurring effect is observed near the equator line ($k_y+A_0(t_0) \sim 0$) advancing toward the poles ($k_x+A_0(t_0) \sim 0$). 
This blurring effect is due to the angular streaking motion of the distributions with time delay, i.e., as the XUV pump pulse acts during a certain time ($T/6$ in our simulations), the PMD moves in the $\hat{x}-\hat{y}$ plane to the left (see Fig. \ref{fig5:kz0atAvstime}a), consequently, the edges of the PMD are not sharp. 
As the width of the PMD is different for different angles in the $(k_x,k_y,0)$ plane, the blurring effect is more pronounced near the equator because the width of the PMD in the direction of angular streaking motion is narrower than the width near the poles. 
In fact, this blurring effect is analogous to normal photography in which the shutter time is comparable to the motion of the snapped object as Louis DiMauro showed more than twenty years ago in the dawn of attophysics \cite{DiMauro02}. 
In Fig. \ref{fig6:kz0.Morecases} we also observe the Cooper-like minimum as the region when the PMD intersects the XUV absorption nodal line $k^2=v_0^2$ (orange line).

%%%%%%%%%%%%%%%%%%%%%%%%%%%%

%-------------FIG 7--------------------------------------
%-------------FIG 8--------------------------------------
%------------------------------------------------------

In Figs. \ref{fig7:ky0.MoreCases1} and \ref{fig8:ky0.MoreCases2}, we present the PMD in momentum space for \(k_y = 0\), analogous to the representation in Fig.~\ref{fig3:SCM}. 
The emission times are set to \(t_0 = 0\) and \(t_0 = T/4\), corresponding to the respective green and red curves in the SCM regions shown in Fig.~\ref{fig3:SCM}.
The parameter sets A through F from Table~\ref{T1} were used for the subplots with matching labels. 
We observe almost circular contours consistent with the predictions of the SCM condition given by Eq.~(\ref{dsdt0}). 
Notably, the PMD exhibits a minimum value for \(k_z \simeq 0\) across all parameter set. This minimum is not strictly zero, as it would be in the dipole approximation, but rather a small nondipole contribution of order \(|1/c|^2\). 
Conversely, a maximum value is observed for \(k_x \simeq 0\) at \(t_0 = 0\) across all parameter sets, which is consistently wider and smaller than the PMD for \(t_0 = T/4\). 
This occurs because, at \(t_0 = 0\), the center of the SCM surface lies outside the \(k_y = 0\) plane. 
The distance to the center, \(|A_0|=F_{L0}/\sqrt{2}\omega\), increases taking the values \(\simeq 0.71\), \(1\), \( 1.7\), \( 2.45\), \(2.8\), and \(4.9\) for cases A through F, respectively. 
However, any contributions near the threshold (\(k = 0\)) would be obscured by the direct ionization term stemming from the IR pulse, which is not considered here.
The maxima in the PMD for \(t_0 = T/4\) (red regions) do not align exactly with the north and south poles of the circumference (as appeared to do in case A), but they are consistently displaced to the right. 
This displacement is particularly evident in case E. 
Furthermore, there is no reflection symmetry with respect to the \(k_z = 0\) plane, as discussed in the SCM representation shown in Fig.~\ref{fig3:SCM}. 
The exact geometry predicted by the SCM will be addressed in a forthcoming publication. 
In the zoom of Fig. \ref{fig8:ky0.MoreCases2}, corresponding to case F, two important features can be observed. 
First, the spectrum deviates significantly from the circle predicted by the SCM under the dipole approximation. Second, the spectrum displays a complex interference pattern, which does not arise from the previously discussed interpulse interference, i.e., they are not sidebands, since Fig.~\ref{fig8:ky0.MoreCases2} corresponds to the streaking PMD without the interpulse factor. 
This interference pattern stems from the Fourier transform of the XUV profile given by its envelope function $F_{X0}(t)$ of duration $\tau_X$ in Eq.~(\ref{envelopeX}).

%---------------------------------
\section{Conclusions and Outlook} \label{conclusions}
%---------------------------------

This work presents a theoretical model to describe circularly polarized laser-assisted photoelectron emission processes beyond the dipole approximation.
Specifically, we focus on the train pulse regime, where periodic properties allow the photoelectron momentum distribution (PMD) to be expressed as a product of the interpulse sideband contribution and the streaking PMD. 
To study a concrete case, we consider the ionization of a hydrogen atom initially in its $1s$ state and systematically explore the transition from the dipole to the nondipole regime by varying the laser parameters.

Our key findings are as follows:
(i) The PMD is predominantly distributed on a surface predicted by the semiclassical model (SCM) in momentum space. 
This is a spherical surface in the dipole regime but gradually deforms as the nondipole parameter $\beta_0$ increases, breaking the forward-backward symmetry.
(ii) The PMD exhibits a non-zero contribution in the plane $k_z=0$, which is a forbidden region within the dipole approximation. 
In this plane, a nodal line resulting from a Cooper-like minimum is observed, and this feature is independent of the polarization state of the infrared (IR) laser.
(iii) The PMD exhibits major signal in the direction opposite to the instantaneous polarization vector. 
Adjusting the time delay $t_0$, the PMD can provide useful information for attoclock chronoscopy.
A particularly significant result is that the interpulse sidebands shift in the direction opposite to the IR propagation, whereas the streaking pattern shifts in the direction opposite to the instantaneous polarization vector. 
Since the polarization and propagation directions are orthogonal in a propagating electromagnetic wave, these shifts occur in mutually orthogonal planes.

In summary, our investigations of nondipole laser-assisted photoemission (LAPE) in both the ATP and streaking regimes provide new insights into time-resolved photoelectron patterns. 
These findings could assist the experimental community in designing time-resolved measurement schemes to observe the transition from dipole to nondipole behavior.
It would be desirable experimental validation of our theoretical predictions for circularly polarized IR lasers.

%---------------------------------
\section{Acknowledgments} 
%---------------------------------

This work was supported by Grants No. PICT-2020-01755, No. PICT-2020-01434, and No. PICT-2017-2945 from ANPCyT (Argentina) and Grant No. PIP 11220210100468CO from CONICET (Argentina).
S. D. L. acknowledges support by Programa Nacional RAICES Federal: Edición 2022 of MinCyT (Argentina) and PICT 2023 RAÍCES FEDERAL 01-PICT-2023-02-10 from ANPCyT (Argentina).
M. F. C. acknowledges support by the National Key Research and Development Program of China (Grant No.~2023YFA1407100), Guangdong Province Science and Technology Major Project (Future functional materials under extreme conditions - 2021B0301030005) and the Guangdong Natural Science Foundation (General Program project No. 2023A1515010871).

%---------------------------------
\bibliography{biblio}% Produces the bibliography via BibTeX.
%---------------------------------

\end{document}